\documentclass[twoside]{dis07}
\usepackage[latin1]{inputenc}
\usepackage[dvips]{graphicx,epsfig,color}
\usepackage{wrapfig,rotating}
\usepackage{amssymb,amsmath,array}

\usepackage{url}

\begin{document}
\title{\textsc{HERMES} Measurements of Collins and Sivers
    Asymmetries from a transversely polarised Hydrogen Target} 

%***********************************************************************
% AUTHORS INFORMATION AREA
%***********************************************************************
\author{Markus Diefenthaler (on behalf of the \textsc{HERMES} collaboration)\\
%
% Optional short acknowledgement: remove next line if non-needed
%\thanks{This is an optional funding source acknowledgement.}
%
% DO NOT MODIFY THE FOLLOWING '\vspace' ARGUMENT
\vspace{.3cm}\\
%
% Addresses and institutions (remove "1- " in case of a single institution)
Friedrich-Alexander-Universit\"at Erlangen-N\"urnberg - Physikalisches
Institut II\\
Erwin-Rommel-Stra{\ss}e 1, 91058 Erlangen - Germany}
%
% Remove the next three lines in case of a single institution
%\vspace{.1cm}\\
%2- School of Second Author - Dept of Second Author \\
%Address of Second Author's school - Country of Second Author's school\\
%}
%***********************************************************************
% END OF AUTHORS INFORMATION AREA
%***********************************************************************

\maketitle

\begin{abstract}
Azimuthal single-spin asymmetries (\textsc{SSA}) in semi-inclusive
electroproduction of \(\pi\)-mesons and charged \(K\)-mesons in
deep-inelastic scattering of positrons and electrons on a transversely
polarised hydrogen target were observed. Significant \textsc{SSA}
amplitudes for both the Collins and the Sivers mechanism are presented
for the full data set recorded with transverse target polarisation at
the \textsc{HERMES} experiment.
\end{abstract}

\section{Contribution} 

In 2005 the \textsc{HERMES} collaboration published first evidence for
azimuthal single-spin asymmetries (\textsc{SSA}) in the semi-inclusive
production of charged pions on a transversely polarised hydrogen
target \cite{Airapetian:2004tw}. Significant signals for both the
Collins \cite{Collins:1992kk} and Sivers mechanisms
\cite{Sivers:1989cc} were observed in data recorded during the
2002--2003 running period of the \textsc{HERMES} experiment.  Below we
present a preliminary analysis of these data combined with additional
data taken in the years 2003--2005; i.e.~an preliminary analysis of
the full data set with transverse target polarisation
\cite{DISTalk:2007}. All data were recorded at a beam energy of
\(27.6\,\text{\texttt{GeV}}\) using a transversely nuclear-polarised
hydrogen-target internal to the \textsc{HERA} storage ring at
\textsc{DESY}. The \textsc{HERMES} dual-radiator ring-imaging
\v{C}erenkov counter allows full \(\pi^{\pm}\), \(K^{\pm}\), \(p\)
separation for all particle momenta within the range
\(2\,\texttt{GeV}<\boldsymbol{P}_\text{h}<15\,\texttt{GeV}\). Therefore,
a preliminary analysis of \textsc{SSA} in the electroproduction of
charged kaons on a transversely polarised target is also presented. In
addition the measurement is accompanied by an preliminary analysis of
reconstructed neutral-pion events.

At leading twist, the longitudinal momentum and spin of the quarks
inside the nucleon are described by three parton distribution
functions: the well-known momentum distribution
\(q\left(x,Q^{\,2}\right)\), the known helicity distribution
\(\Delta\,q\left(x,Q^{\,2}\right)\) \cite{Airapetian:2004zf} and the
(experimentally) unknown transversity distribution
\(\delta\,q\left(x,Q^{\,2}\right)\) \cite{Ralston:1979ys,
Artru:1989zv, Jaffe:1991kp, Cortes:1991ja}. In the helicity basis,
transversity is related to a quark-nucleon forward scattering
amplitude involving helicity flip of both nucleon and quark
(\(N^{\Rightarrow}q^{\leftarrow} \boldsymbol{\rightarrow}
N^{\Leftarrow}q^{\rightarrow}\)). As it is chiral-odd, transversity
cannot be probed in inclusive deep-inelastic scattering
(\textsc{DIS}). At \textsc{HERMES} transversity in conjunction with
the chiral-odd Collins fragmentation function \cite{Collins:1992kk} is
accessible in \textsc{SSA} in semi-inclusive \textsc{DIS} on a
transversely polarised target (\emph{Collins mechanism}). The Collins
fragmentation function describes the correlation between the
transverse polarisation of the struck quark and the transverse
momentum \(\boldsymbol{P}_{\,\text{h}\perp}\) of the hadron
produced. As it is also odd under naive time reversal (T-odd) it can
produce a \textsc{SSA}, i.e.~a left-right asymmetry in the momentum
distribution of the produced hadrons in the plane transverse to the
virtual photon direction.

The \emph{Sivers mechanism} can also cause a \textsc{SSA}: The T-odd
Sivers distribution function \cite{Sivers:1989cc} describes the
correlation between the transverse polarisation of the nucleon and the
transverse momentum \(\boldsymbol{p}_{\perp}\) of the quarks within. A
non-zero Sivers mechanism provides a non-zero Compton amplitude
involving nucleon helicity flip without quark helicity flip
(\(N^{\Rightarrow}q^{\leftarrow} \boldsymbol{\rightarrow}
N^{\Leftarrow}q^{\leftarrow}\)), which must therefore involve orbital
angular momentum of the quark inside the nucleon
\cite{Burkardt:2003yg, Brodsky:2002cx}.

\begin{figure}[t]
\begin{tabular}{cc}
\includegraphics[scale=0.33]{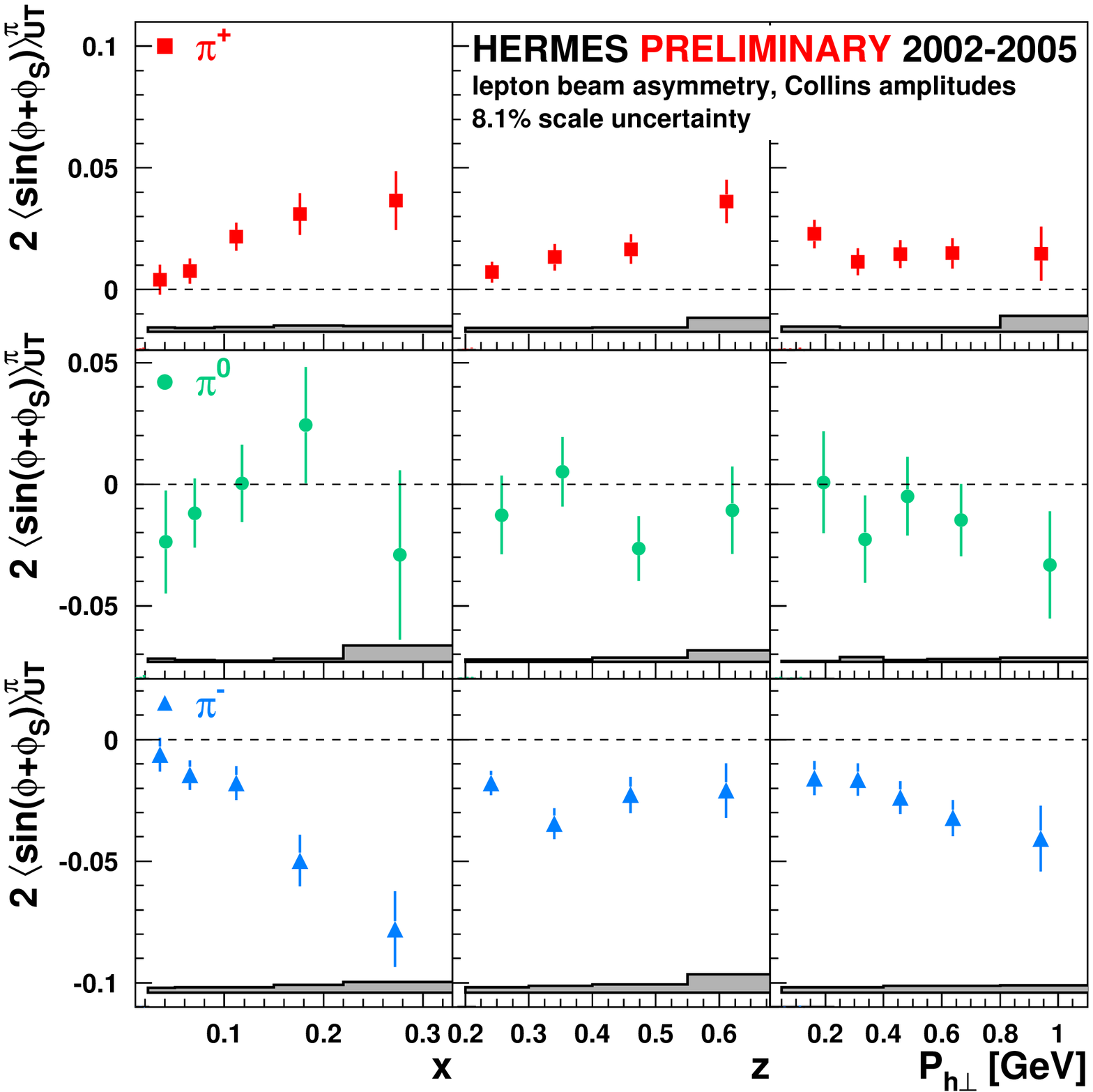} &
\includegraphics[scale=0.33]{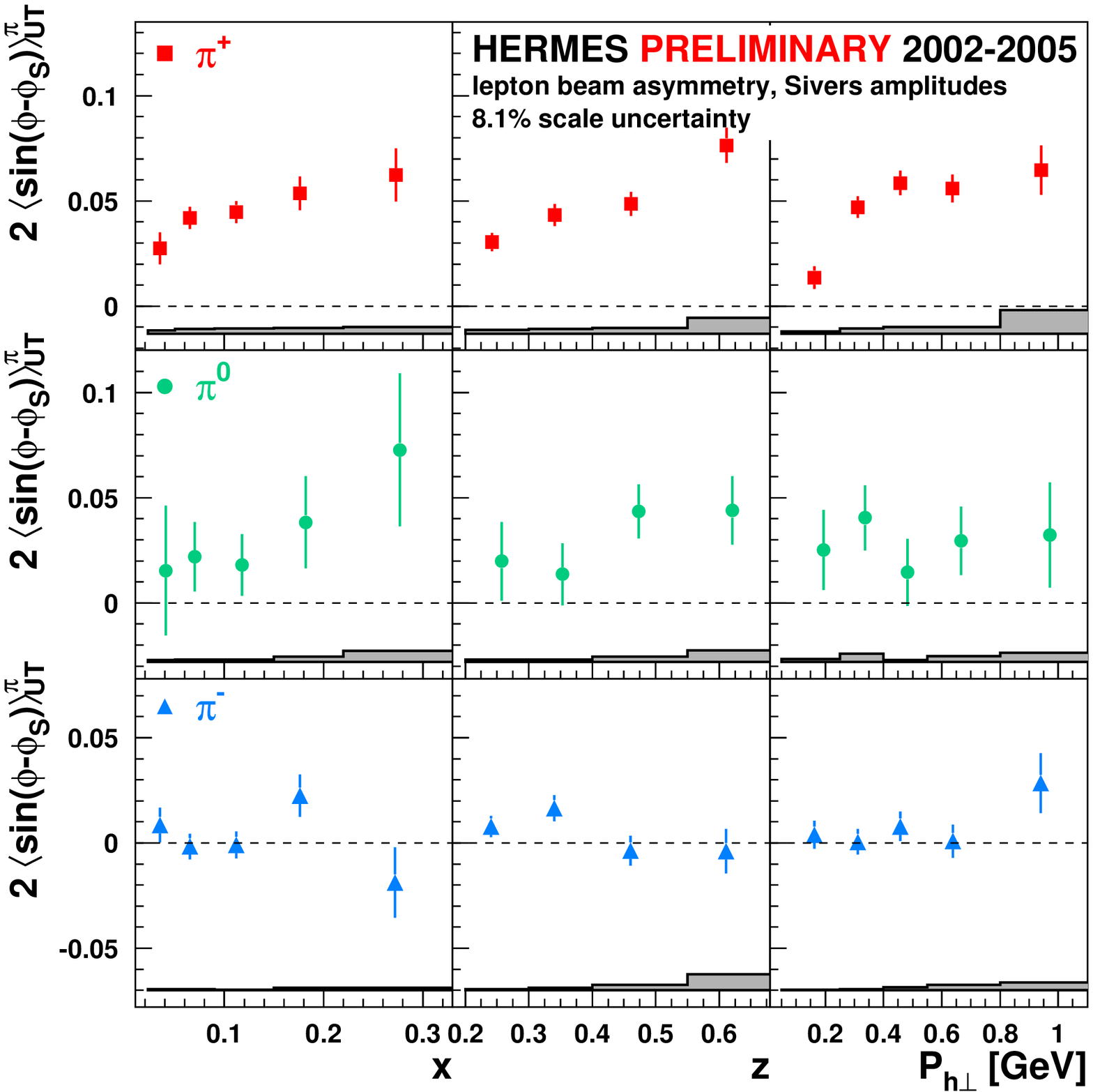}
\end{tabular}
\caption{Collins amplitudes (left column) and Sivers amplitudes (right
column) for \(\pi\)-mesons (as labelled) as function of \(x\), \(z\)
and \(\boldsymbol{P}_{\,\text{h}\perp}\). The error bands represent
the maximal systematic uncertainty; the common overall
\(8.1\%\) scaling uncertainty is due to the target polarisation
uncertainty.}
\label{ssa-pimesons}
\end{figure}

With a transversely polarised target, the azimuthal angle
\(\phi_{\,\text{S}}\) of the target spin direction in the
``\(\Uparrow\)'' state is observable in addition to the azimuthal
angle \(\phi\) of the detected hadron.  Both azimuthal angles are
defined about the virtual-photon direction with respect to the lepton
scattering plane.  The additional degree of freedom
\(\phi_{\,\text{S}}\), not available with a longitudinally polarised
target, results in distinctive signatures: \(\sin{\left(\phi +
\phi_{\,\text{S}}\right)}\) for the Collins mechanisms and
\(\sin{\left(\phi - \phi_{\,\text{S}}\right)}\) for the Sivers
mechanism \cite{Boer:1997nt}.

The corresponding azimuthal amplitudes azimuthal amplitudes are
extracted using maximum likelihood fits. The Collins amplitudes
\(\left<\sin{\left(\phi +
\phi_{\,\text{S}}\right)}\right>_{\text{UT}}^\text{h}\) and the Sivers
mechanism \(\left<\sin{\left(\phi -
\phi_{\,\text{S}}\right)}\right>_{\text{UT}}^\text{h}\) were extracted
simultaneously to avoid cross-contamination. To allow for contribution
from all theoretically possible Fourier modulations the terms for
\(\sin{\phi_{\,\text{S}}}\), \(\sin{\left(2\phi-
\phi_{\,\text{S}}\right)}\) and \(\sin{\left(3\phi-
\phi_{\,\text{S}}\right)}\) have to be added in the probability
density function \(\operatorname{F}\):
\begin{eqnarray*}
\operatorname{F}\left(2\left<\sin{\left(\phi\pm\phi_S\right)}\right>_{\text{UT}}^\text{h},\ldots,\phi,\phi_S\right)
& = \cfrac{1}{2}\Big(1+P_{\alpha}^z\big(&2\left<\sin{\left(\phi+\phi_S\right)}\right>_{\text{UT}}^\text{h}\cdot\sin{\left(\phi+\phi_S\right)}+\\
& &2\left<\sin{\left(\phi-\phi_S\right)}\right>_{\text{UT}}^\text{h}\cdot\sin{\left(\phi-\phi_S\right)}+\\
& &2\left<\sin{\left(3\phi-\phi_S\right)}\right>_{\text{UT}}^\text{h}\cdot\sin{\left(3\phi-\phi_S\right)}+\\
& &2\left<\sin{\left(2\phi-\phi_S\right)}\right>_{\text{UT}}^\text{h}\cdot\sin{\left(2\phi-\phi_S\right)}+\\
& &2\left<\sin{\phi_S}\right>_{\text{UT}}^\text{h} \cdot\sin{\phi_S}\big)\Big)\\
\end{eqnarray*}
Here \(P_\alpha^z\) denotes the degree of the target polarisation.

\begin{figure}
\begin{tabular}{cc}
\includegraphics[scale=0.33]{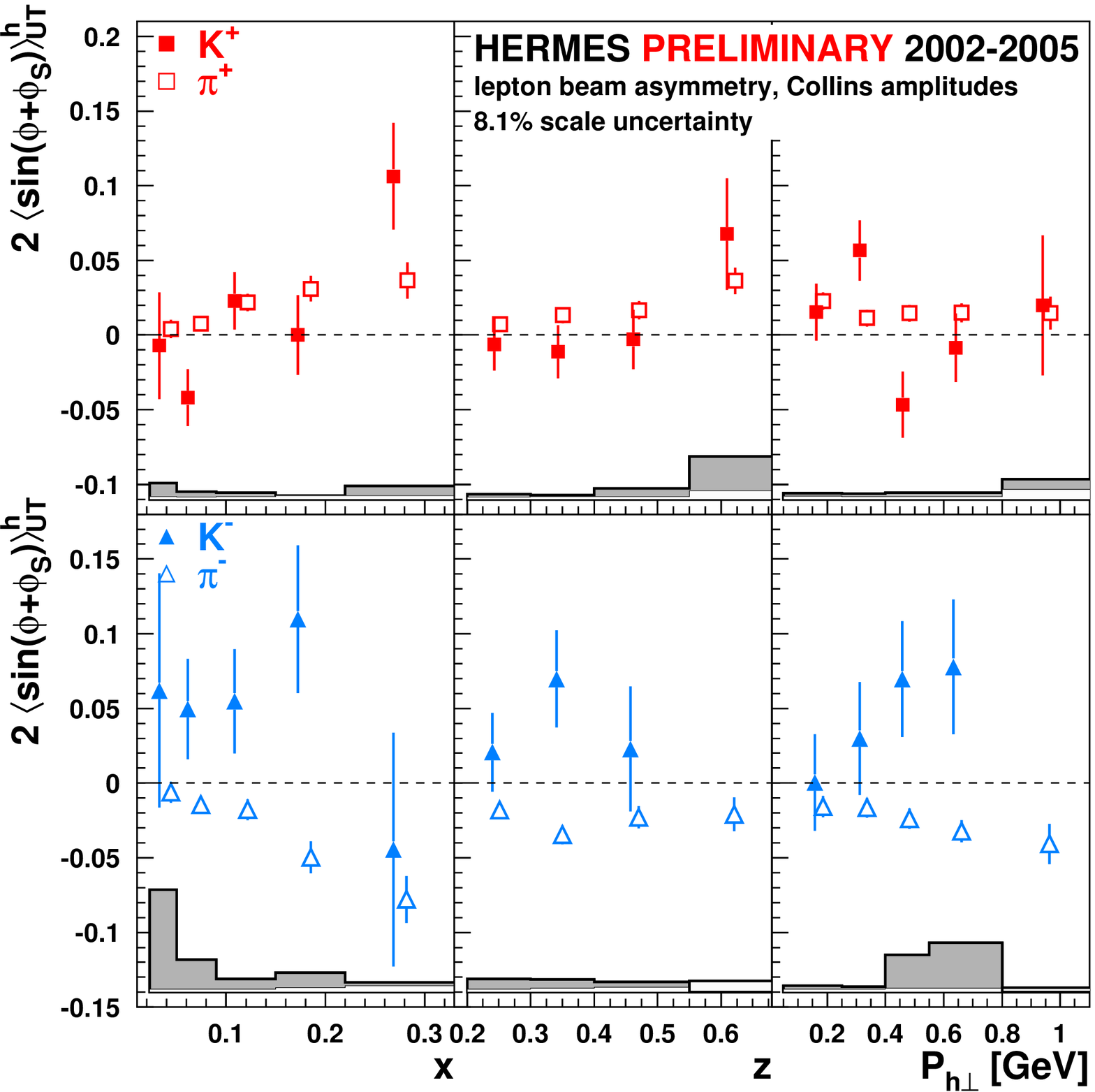} &
\includegraphics[scale=0.33]{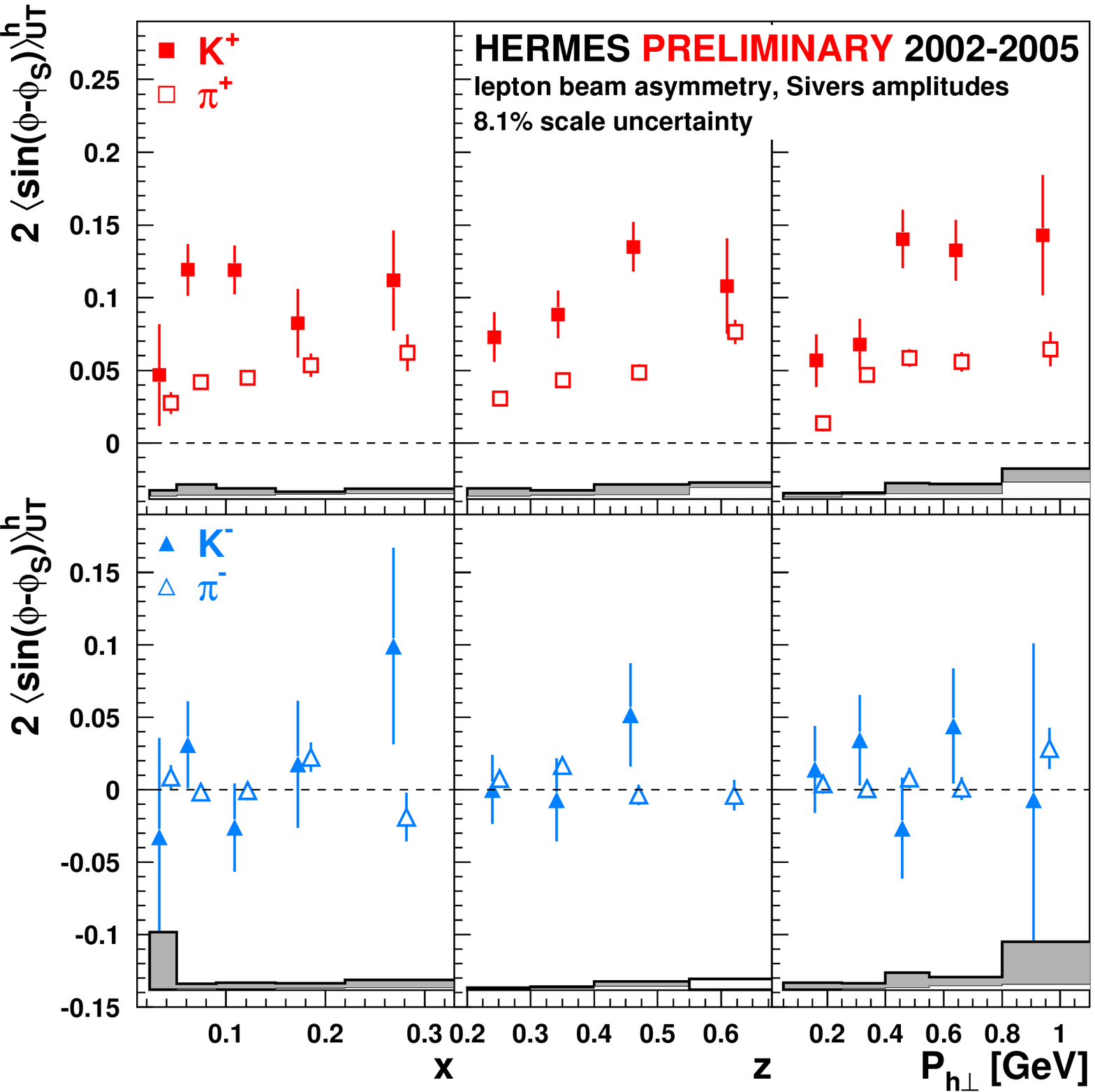}
\end{tabular}
\caption{Collins amplitudes (left column) and Sivers amplitudes (right
column) for charged kaons (closed symbols, as labelled) and charged
pions (open symbols, as labelled) as function of \(x\), \(z\)
and \(\boldsymbol{P}_{\,\text{h}\perp}\). The error bands represent
the maximal systematic uncertainty; the common overall
\(8.1\%\) scaling uncertainty is due to the target polarisation
uncertainty.}
\label{ssa-mesoncomparion}
\end{figure}

In Figures \ref{ssa-pimesons} and \ref{ssa-mesoncomparion} the
lepton-beam Collins and Sivers amplitudes as a function of \(x\),
\(z\) and \(\boldsymbol{P}_{\,\text{h}\perp}\) are
shown. Semi-inclusive \textsc{DIS} events were selected subject to the
kinematic requirements \(Q^2>1\,\text{\texttt{GeV}}^2\), \(y<0.95\),
\(W^2>10\,\text{\texttt{GeV}}^2\),
\(2\,\texttt{GeV}<\boldsymbol{P}_\text{h}<15\,\texttt{GeV}\),
\(0.2<z<0.7\) and
\(\theta_{\gamma^*\text{h}}>0.02\,\text{\texttt{rad}}\), where
\(\theta_{\gamma^*\text{h}}\) is the angle between the direction of
the virtual photon and the hadron. The selected ranges in \(x\) and
\(\boldsymbol{P}_{\,\text{h}\perp}\) are \(0.023<x<0.4\) and
\(0.05\texttt{GeV}<\boldsymbol{P}_{\,\text{h}\perp}<2\texttt{GeV}\). These
preliminary results are based on ten times more statistics than that
in the publication \cite{Airapetian:2004tw} and are consistent with
the published result.

The Collins amplitude is positive for \(\pi^+\), compatible with zero
for \(\pi^0\) and negative for \(\pi^-\). Also, the magnitude of the
\(\pi^-\) amplitude appears to be comparable or larger than the one
for \(\pi^+\). This leads to the conclusion that the disfavoured
Collins fragmentation function has a substantial magnitude with an
opposite sign compared to the favoured Collins fragmentation
function. For charged kaons no statistically significant non-zero
Collins amplitudes are found. However, the Collins amplitudes for
\(K^+\) are within statistical accuracy consistent to the \(\pi^+\)
amplitudes.

The significantly positive average Sivers amplitudes observed for
\(\pi^+\), \(\pi^0\) and \(K^+\) imply a non-vanishing orbital angular
momentum of the quarks inside the nucleon. As the magnitude of the
\(K^+\) amplitude is the larger than the one for \(\pi^+\), the sea
quark contribution to the Sivers mechanism appears to be
important. Thus the orbital angular momentum of anti-quarks could be
significant and highly flavour dependent. For \(\pi^-\) and \(K^-\)
the Sivers amplitudes are consistent with zero.

Isospin symmetry of \(\pi\)-mesons is fulfilled for the extracted
Collins and Sivers amplitudes.

\begin{figure}[t]
\begin{tabular}{cc}
\includegraphics[scale=0.33]{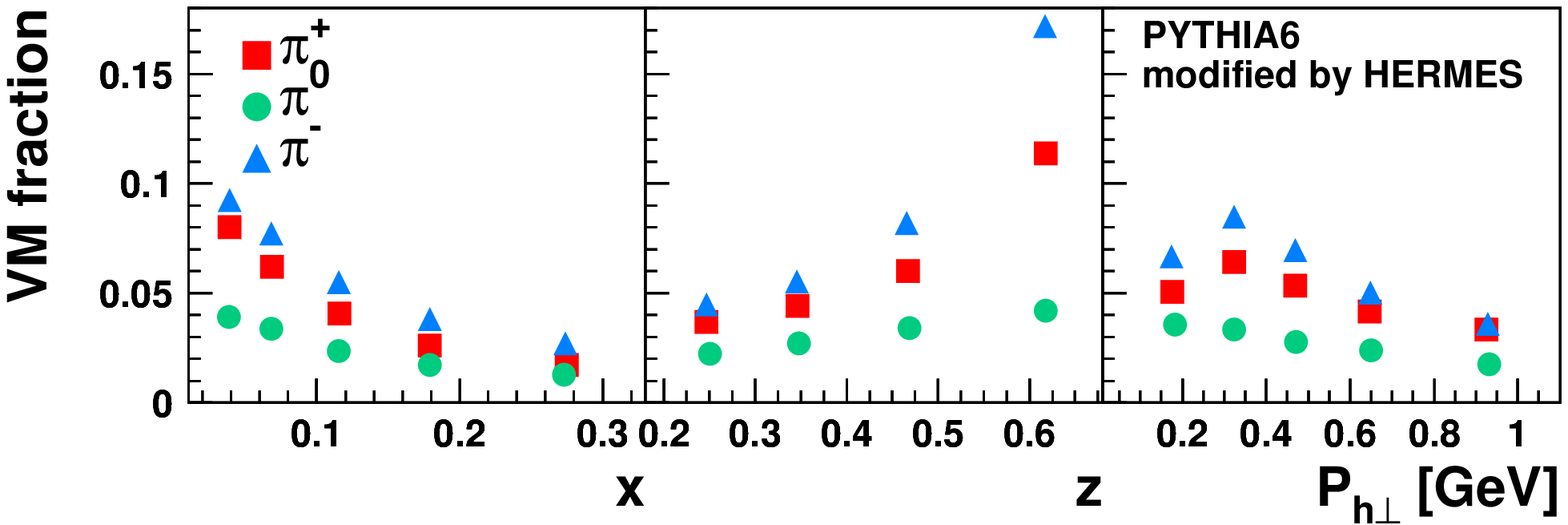} &
\includegraphics[scale=0.33]{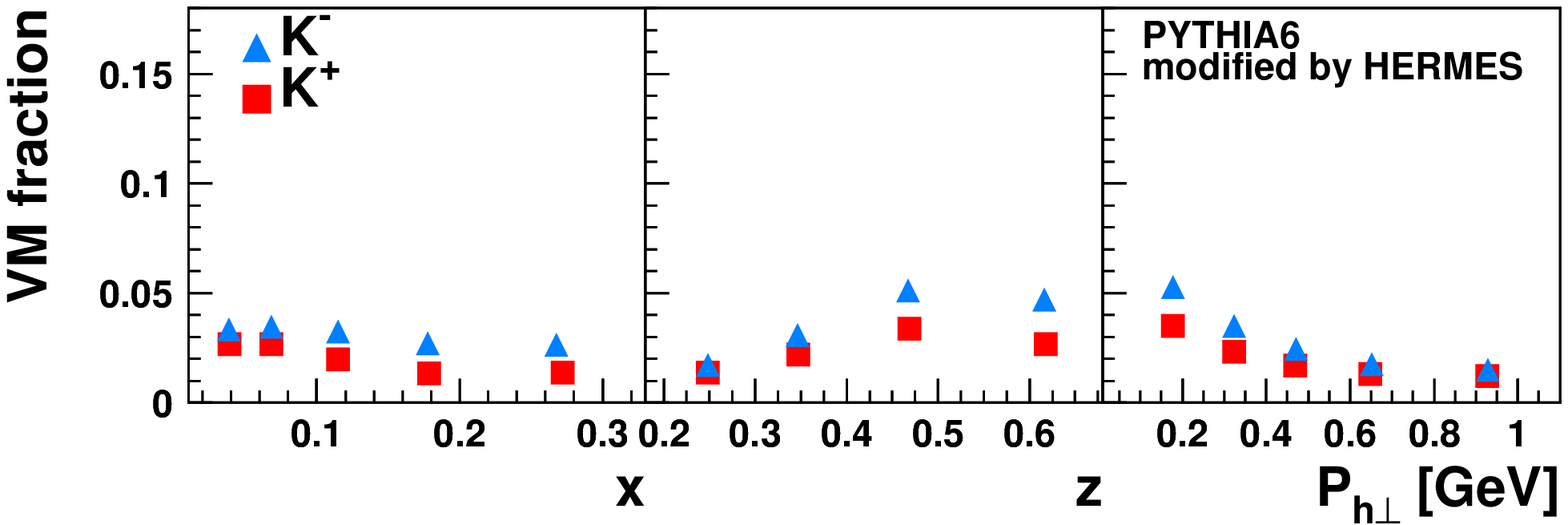}\\
\end{tabular}
\caption{Simulated fraction of \(\pi\)-mesons (left column) and
  charged \(K\)-mesons originating from diffractive vector meson
  production and decay.}
\label{ssa-vmdilution}
\end{figure}

Although formally, the contribution to the \textsc{SSA} of
\(\pi\)-mesons and \(K\)-mesons from the decay of exclusively produced
vector mesons is a part of the semi-inclusive \textsc{DIS} cross
section, a too large contribution might contradict the assumptions of
factorisation, i.e.~summation over a larger number of contributing
channels. As an indication, the simulated fraction of \(\pi\)-mesons
and charged \(K\)-mesons originating from diffractive vector meson
production and decay is shown in Figure \ref{ssa-vmdilution}

\section{Acknowledgements}

This work has been supported by the German Bundesministerium f\"ur
Bildung und Forschung (\textsc{BMBF}) (contract nr.~06 \textsc{ER}
125I and 06 \textsc{ER} 143) and the European Community-Research
Infrastructure Activity under the FP6 ''Structuring the European
Research Area'' program (HadronPhysics I3, contract
nr.~RII3-CT-2004-506078).

\section{Bibliography}
%To offer a link to the slides of your contribution we ask you to 
%provide the URL as first reference. An example is given below~\cite{url}.
%You have to replace the numbers for \texttt{contribId}   and   \texttt{sessionId}
%(42 and 8, respectively in the example).

%You can get this information by going to \\
%\verb$http://indico.cern.ch/confAuthorIndex.py?confId=9499$,
%search for your contribution, click on the title and take the information
%from the listed URL. 

%Be aware: \verb$&amp;$ must be replaced by simple \verb$&$ as in the example.
%dvipdf will include an active link to the pdf file.
 
% ****************************************************************************
% BIBLIOGRAPHY AREA
% ****************************************************************************

\begin{footnotesize}

\end{footnotesize}

% ****************************************************************************
% END OF BIBLIOGRAPHY AREA
% ****************************************************************************

\end{document}